\documentclass[conference]{IEEEtran}
\IEEEoverridecommandlockouts

\usepackage{amsmath,amssymb,amsfonts}
\usepackage{algorithmic}
\usepackage{graphicx}
\usepackage{textcomp}
\usepackage{xcolor}
\def\BibTeX{{\rm B\kern-.05em{\sc i\kern-.025em b}\kern-.08em
    T\kern-.1667em\lower.7ex\hbox{E}\kern-.125emX}}

\usepackage{float}   
\usepackage{authblk}     
\usepackage[backend=biber,sorting=none]{biblatex}
\begin{document}

\title{\LARGE \bf helyOS: A customized off-the-shelf solution for autonomous driving applications in delimited areas} 

\author{Carlos Viol Barbosa$^{*}$,  Nikolay Belov, Felix Keppler,  Julius Kolb, Gunter Nitzsche  and Sebastian Wagner

\thanks{$^{*}${\tt\small violbarbosa@ivi.fraunhofer.de}}
}
 
\affil{Fraunhofer Institute for Transportation and Infrastructure Systems IVI, Zeunerstraße 38, 01069 Dresden, Germany}

\maketitle

\begin{abstract}
Microservice Architectures (MSA), known to successfully handle complex software systems, are emerging as the new paradigm for automotive software. The design of an MSA requires  correct subdivision of the software system and implementation of the communication between components. These tasks demand both software expertise and domain knowledge.  In this context, we developed an MSA framework pre-tailored to meet the requirements of autonomous driving applications in delimited areas - the helyOS framework. The framework decomposes complex applications in predefined microservice domains and provides a communication backbone for event messages and data. This paper demonstrates how such a tailored MSA framework can accelerate the development by prompting a quick start for the integration of motion planning algorithms, device controllers, vehicles simulators and web-browser interfaces.  
\end{abstract}


\section{Introduction}
Autonomous driving and fleet management are complex applications  at the intersection of multiple technologies. These fields have become important within the last years and prompted many researches focused on  motion planners ~\cite{planning, trucktrix_path}, cooperative driving~\cite{trucktrix_coop}, platoon formation control~\cite{robotics_control}, 3D mapping~\cite{3d_mapping} and artificial intelligence~\cite{deep_learning}. A promising sub-field of autonomous driving, which can already lead to functional implementations today, is the automation of vehicles (AV) in delimited areas. By delimited area we refer to areas closed for general traffic where an automated vehicle can only meet other automated vehicles or vehicles operated by instructed personnel. This concept can be applied in sea harbors, airports, factories, logistics centers, agricultural fields ~\cite{fleet_book_1}, etc. A common feature in those applications is the existence of tasks or missions to be executed by automated vehicles.

\begin{figure}
\centering
\includegraphics[width=0.4\textwidth]{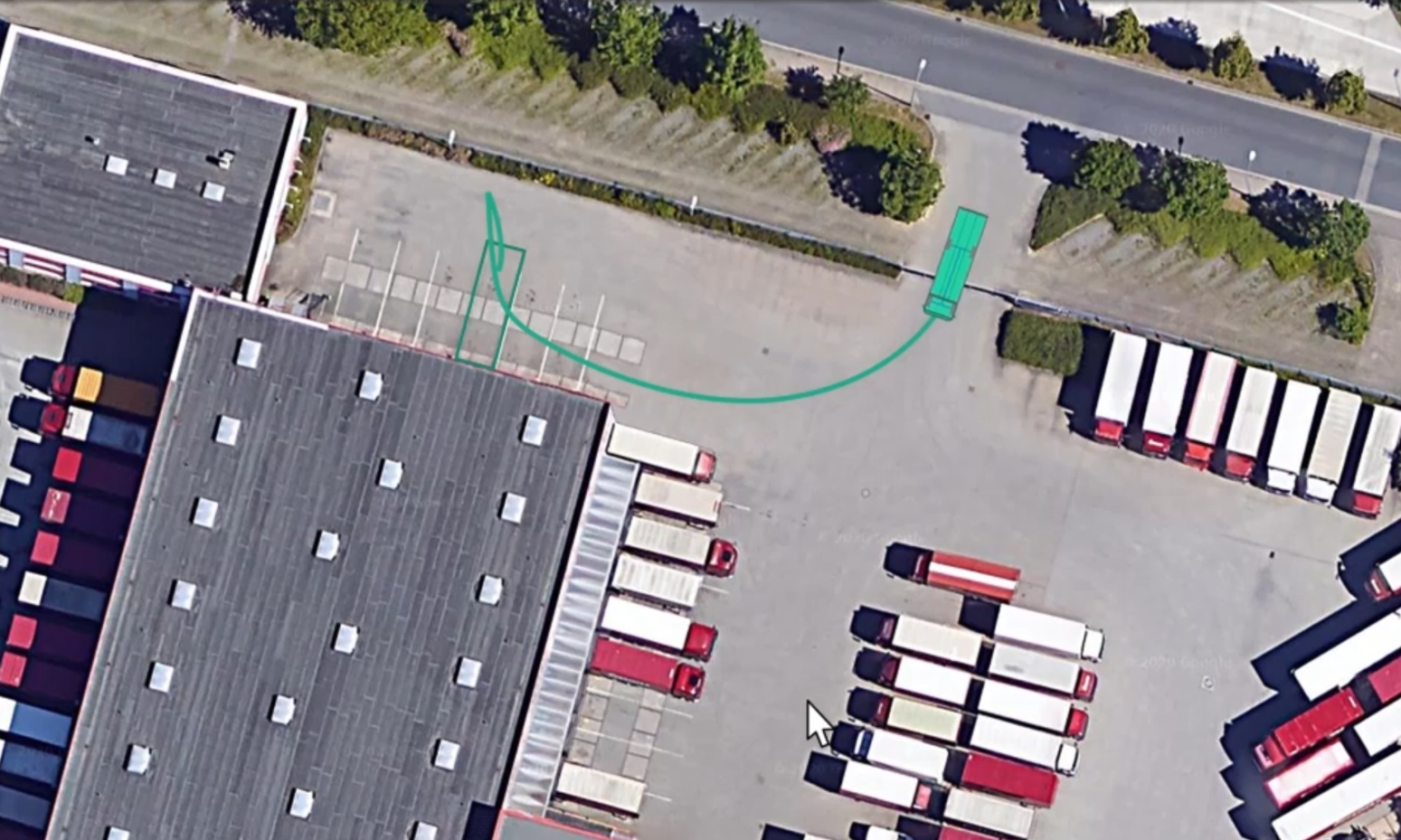}
\caption{Logistics yard in the AutoTruck project.} 
\label{fig:logistic_yard}
\end{figure}

\begin{figure*}
\centering
\includegraphics[width=\textwidth]{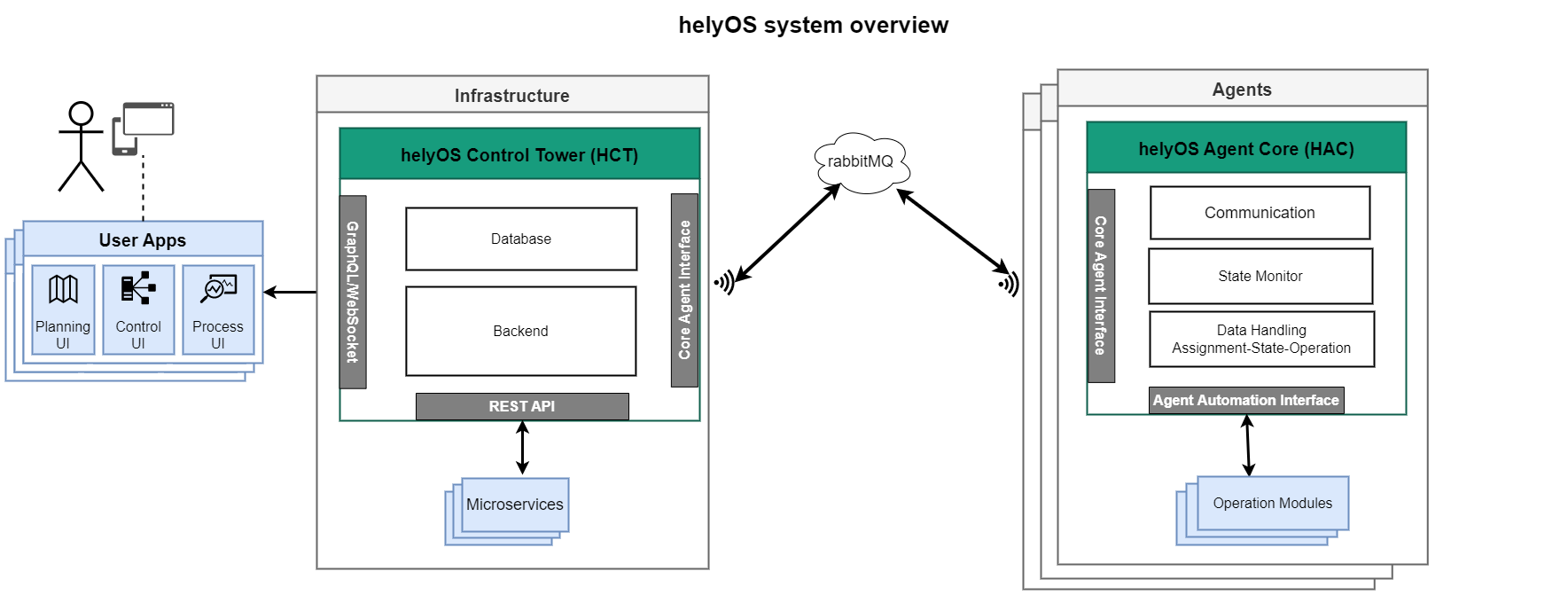}
\caption{helyOS system overview. The helyOS Control Tower (HCT) is simultaneously used as backend for \textbf{User Apps}, as microservice controller to coordinate external services, and as a mission dispatcher to cast assignments to \textbf{Agents}. 
The helyOS Agent Core (HAC) handles the execution of the mission assignments by the automated tool.
helyOS is an event-driven architecture using  HTTP, WebSocket and AMQP protocols.  }
\label{fig:helyos}
\end{figure*}

Combining different technologies to enable automatons to perform missions is challenging~\cite{fleet_book_2}. Most frequently, such projects deal with numerous partners from different domains; decisions about how to interface and coordinate different software and hardware systems consume a great deal of time. Microservice architecture (MSA) has been suggested~\cite{Hoffmann_2017, MSA_2019}  as a strategy to overcome such integration hassles. MSA has been successfully applied to create flexible, maintainable, and scalable information systems. Regarding automotive software systems, MSA has several advantages~\cite{MSA_2019}: (i) reuse of functionality, (ii) encapsulated and independent service behavior, (iii) continuous service integration, (iv) hierarchical in-vehicle function and software architecture,  (v) flexible service distribution to electronic control units (ECUs) or cloud servers, (vi) modular software design is easier to understand than big opaque monoliths.

The use of microservices, however, has the potential risks of adding too much granularity to the application and overlaps between domain boundaries, which hinder system maintenance. Avoiding these pitfalls demands knowledge and experience from the developers. In this context, we developed helyOS (highly efficient online yard Operating System) as an MSA framework designed for autonomous driving in delimited areas. Our approach mitigates the MSA risks by providing a framework with pre-defined data flow rules and boundaries between common domains in AV: mission data processing, task assignment control, mission execution, geo-map  processing and user interface.

In association with rabbitMQ \cite{RabbitMQ}, helyOS  also provides a communication backbone for exchanging data and events. The registration of the system entities (microservices, automatons, maps and mission descriptions) is performed via a graphic admin interface. The system is therefore easy tailored to a desired application and most of the development efforts can be focused on the automatons and end-user features.

In the present paper, we firstly give some background information and an overview of the helyOS architecture
 (Sec.~\ref{sec:background} and ~\ref{sec:architecture}), then describe how helyOS handles different microservices to compose and distribute the assignments to the robots   (Sec.~\ref{sec:hct}). Finally, we present several scenarios of yard automation covered by helyOS (Sec.~\ref{sec:scenarios}). Furthermore, in our conclusion (Sec.~\ref{sec:concl}), we list ongoing projects that are already making use of the helyOS architecture.

\section{Background}\label{sec:background}

 The employment of modular monolithic systems is a common practice in automotive software. Modular systems are certainly able to comply with domain-driven-design (DDD) concepts \cite{evans2004ddd}, however, the final integration is strongly conditioned by the linking or compiling of each code package into the application deployment. In large projects, where professionals of different technical backgrounds and organizations must collaborate, the solely use of a modular monolithic approaches will impose restrictions on the development.
 
 Microservices, on the other hand, allow software integration independent of programming language or hardware. Since gateways for web applications and cloud computation can be easily defined, MSA provides outsourcing opportunities not present in monolithic modular systems. Indeed, helyOS originated from the necessity of employing computationally demanding motion planning algorithms running in the cloud for the automation of a logistics yard (Fig. ~\ref{fig:logistic_yard}) within the AutoTruck project \cite{autotruck}. In combination with Docker containerization, the microservice approach has improved the development workflow thanks to a clearer separation of concerns and the interoperability of software components independent of computer platforms. Since then, the helyOS framework has been shaped throughout several projects to be applicable to a broader class of automation problems in which mobile robots perform actions assigned by a central control tower.

\begin{figure*}
\centering
\includegraphics[width=0.8\textwidth]{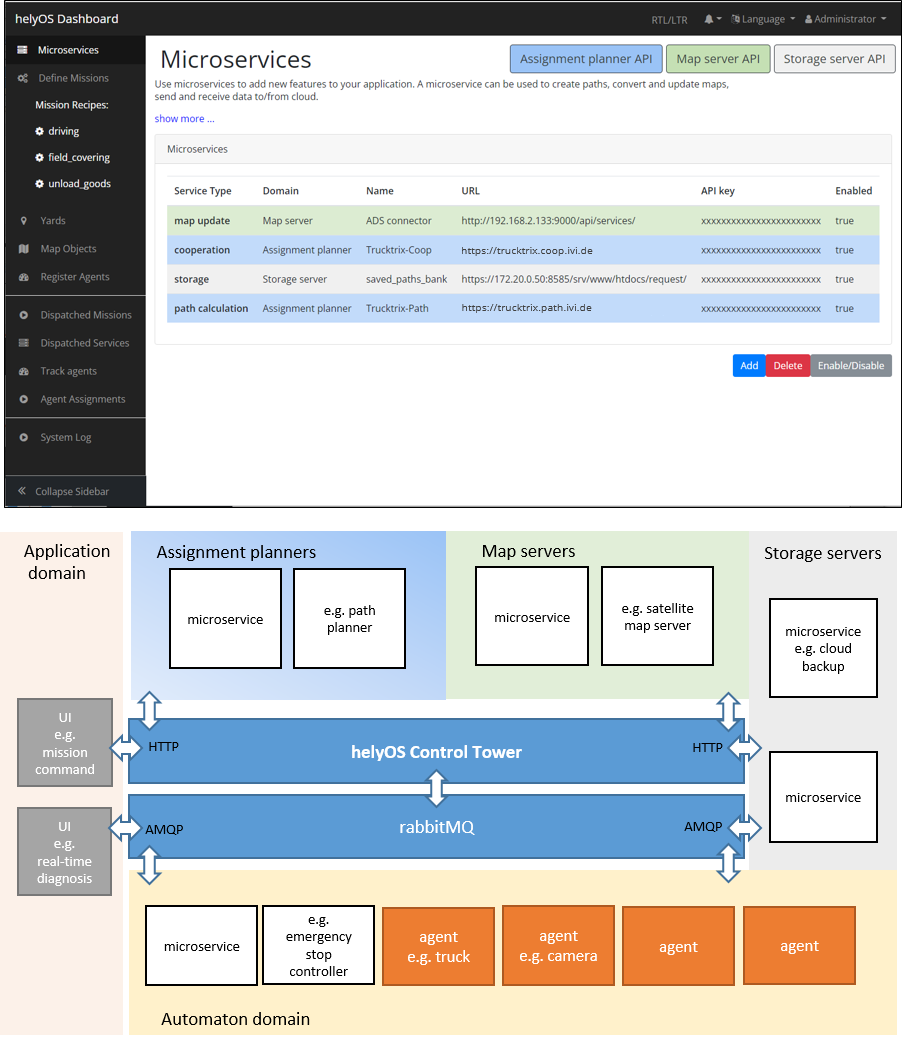}
\caption{Microservices in the helyOS framework: The upper panel shows the helyOS dashboard, where the microservices are registered. Note that each microservice is registered under a specific software domains (see column \textbf{Domain}). The bottom panel illustrates the connection of the microservices of different domains and the HCT. The \textit{Assignment planners} microservices process the request data to create assignments. The microservices in \textit{Map servers} are used to updated the geo-map information in the helyOS database. In \textit{Storage servers}, the microservices are employed to export/backup the yard data. Finally, the \textit{Automaton domain} includes all microservices directly accessible by the agents and independent from the helyOS control tower.}
\label{fig:dashboard_msa}
\end{figure*}

\section{helyOS architecture}
\label{sec:architecture}

An application built on helyOS architecture is composed of two main parts: helyOS Control Tower (HCT)  and helyOS Agent Core (HAC). The HCT is stationary, whereas the HAC is installed in the tools or moving actors: robots, vehicles or other machines. In this manuscript, these tools are referred to as ”agents”.  Both the HCT and the HAC have a general-purpose architecture and project specificities are supplied by microservices and operation modules. The microservices are used for motion planning, cooperative driving, map updates, etc. The operation modules are responsible for the vehicle automation tasks. 

helyOS Control Tower (HCT),  shown on the center panel of Fig.~\ref{fig:helyos}, performs the high-level logic for yard automation. The data flow is driven by the message broker (rabbitMQ) and database events; these events are used to orchestrate the microservices and dispatch assignments. A request for a mission, initiated either by the user app or an automaton, triggers calls of microservices of different functionalities. The response from the microservices will compose the assignment data that will be delivered to the targeted agent via rabbitMQ.

 HCT consists of an internal database and backend.  The database is implemented in Postgres~\cite{pgsql}, an open source object-relational database management system. The backend gives access to the database by employing Postgraphile~\cite{postgraphile}, an implementation of GraphQL~\cite{graphql} that dynamically adapts to changes in the Postgres schema. This means that new columns and tables included in the database are automatically available to the application frontend without extra code in the backend. The HCT backend is implemented in NodeJs and run from a Docker container.

The HCT backend is responsible for the agent registration, agent monitoring, sensor logging, microservices orchestration, assignment dispatching and the storage of the yard state data. It supplies a gateway for user interfaces for sending commands and retrieving data via HTTP and WebSocket. The HCT also provides a convenient graphical interface served locally for configuration purposes; the helyOS dashboard. This means that the system developer/administrator can use a web browser to configure the backend.

 helyOS Agent Core (HAC), shown on the right panel of Fig.~\ref{fig:helyos}, handles the execution of the assignments by the automated tool, which can be a vehicle, a camera, an external sensor or any other actor on the yard. The HAC subscribes to rabbitMQ to receive assignments and to publish its sensor data and status. The sensor data may contain information about geo-position, battery and other diagnosis data, which are stored in the HCT database and broadcast to connected user interfaces. The HAC status defines the current state of the device, and it is continuously published to enable the management and distribution of assignments by the HCT.

Fig.~\ref{fig:helyos} describes helyOS in an application context.  User apps, on the left panel, communicate with helyOS via HTTP requests using the GraphQL query language, while helyOS broadcasts agent sensor data and status notifications by using Web-Socket. In a typical scenario, the user app dispatches a mission by creating a work process instance in the HCT database, this event triggers a number of process requests. After clearing the HAC status, each process request is dispatched to its specific microservice. helyOS organizes the microservice calls and saves the responses in the database. Once all microservice requests are processed, the resulting data is rendered as assignments and delivered to the agents, shown on the right panel, by using the rabbitMQ message broker. It is worthy to mention that the employment of one or several microservices in a mission is preconfigured by the administrator using the helyOS dashboard.

On the ”Agent” side, the HAC receives the assignment from the HCT and splits it up into a set of operation requests. These requests are consumed by dedicated operation modules; each module is responsible for the execution of a specific task. While these task modules may vary across projects and machines, the data flow and the process hierarchy are suggested to be standard in our framework.

\section{ Microservice domains}
\label{sec:hct}

To be accessible to the HCT, a microservice must be registered and ascribed to a domain in the helyOS dashboard, see Fig. ~\ref{fig:dashboard_msa}. The helyOS framework defines four domains for microservices: assignment planners, map servers, storage servers and the automaton domain. Each domain has a specific set of permissions for data flow, which guarantees the consistence of the yard state and guides the overall design of the application. Developers are free to create subdomains if required.

The assignment planners include all microservices intended to process the requests of tasks and operations. Microservices registered in this domain have permission to send data to a targeted agent. These services can be used to create assignments or to simply feed context data to an agent.

Map server microservices have the permission to write map data into the helyOS database. They are used when changes of the map objects are triggered by missions. The map data is saved as a list of JSON objects (user-defined schema) and is part of the yard state data. This data is  relevant for motion algorithms and it is shared with all other microservices.

The storage servers are microservices used to passively output data. Since the yard state data is automatically appended to any request to a microservice, this data can be forwarded to a storage service. Microservices in this domain cannot modify the yard state or send data to agents. Finally, the microservices in the automatons domain work in conjunction with the agents and are independent from helyOS. The autonomous domain is suitable, for example, for the implementation of real-time features such as data streaming between microservices and agents. 

All microservices accessible to HCT must comply with a minimalist REST API for CRUD operations (create, read, update and delete). A microservice can either synchronously return the results, in case of fast processing, or return a job-identifier token (JOB-ID), in case of time-demanding processes. The latter is the typical scenario for motion planning algorithms, which can take several tens of seconds. In the HCT, an HTTP periodic polling mechanism is employed for retrieving the results from the microservice by JOB-ID. This means that the microservices will passively wait to be requested, which keeps their implementation simple. For microservices connected via rabbitMQ, the response ingestion is triggered by event.

\begin{figure*}
\vspace{0.3cm}
\includegraphics[width=\textwidth]{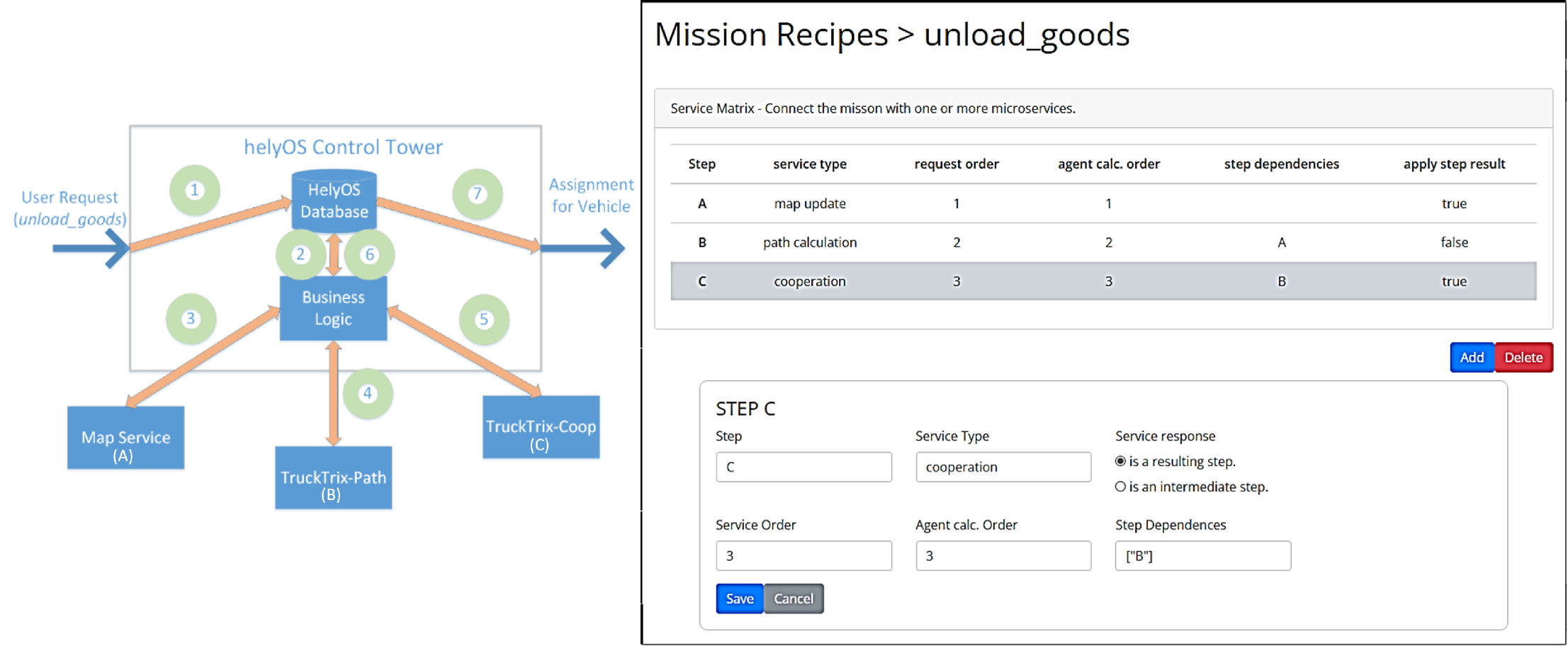}
\caption{The  "unload goods" mission diagram (left panel) and its definition in the helyOS dashboard (right panel). Numbers on the diagram are for: }
\label{fig:mission_recipe}
\end{figure*}

Except for those in the automaton domains, the microservices are triggered by mission requests. A mission does not necessarily have to involve agents; e.g., a mission request to update the yard map will only set an interaction between a microservice and the HCT database. When a mission is being processed, the choice of which microservices are employed must be previously encoded by the administrator/developer as a "mission recipe". Mission recipes can be created and edited in helyOS dashboard, and each new mission recipe translates to a new feature available for the end-user (see right panel Fig. \ref{fig:mission_recipe}).

\section{Examples of scenarios supported by helyOS}
\label{sec:scenarios}

In this section, the following examples of autonomous driving applications make use of motion algorithms from the TruckTrix\textsuperscript{\textregistered} family:
(i) TruckTrix\textsuperscript{\textregistered}-Path~\cite{trucktrix_path} -- path planner for a single vehicle.
(ii) TruckTrix\textsuperscript{\textregistered}-Coop~\cite{trucktrix_coop} --  timestamps  trajectories calculated by TruckTrix\textsuperscript{\textregistered}-Path to avoid collisions.
(iii) TruckTrix\textsuperscript{\textregistered}-Agriculture~\cite{trucktrix_field} -- path planner for efficient weed and pest control using agricultural robots in smart farming.

\subsection{Logistics hub/yard with autonomous vehicles}

Autonomous vehicles in logistics hubs are the flagship application for helyOS. For demonstration purposes, let us consider the following use case:

(i) A truck, carrying goods, arrives at the entrance of the logistics yard.

(ii) The driver sets the truck to autonomous mode, checks it into helyOS and leaves the vehicle.

(iii) The yard personnel use a web app to send the truck to a certain 
\textit{gate} to unload the goods, and then to drive to a \textit{parking slot}, where the truck will be waiting for its next mission.

The AutoTruck project (see Ref.~\cite{autotruck} ) was a practical realization of this scenario using the helyOS framework. Fig.~\ref{fig:mission_recipe} describes how helyOS handles this use case.
In short, the mission request is sent from the web app to the HCT (1) creating a database event (2). This event triggers map updates from a map service (3). The updated map data and truck geometries are then sent to TruckTrix\textsuperscript{\textregistered}-Path (4) for the calculation of a collision-free trajectory from the yard entrance to the \textit{gate}. The result is then forwarded to TruckTrix\textsuperscript{\textregistered}-Coop  (5) that updates the trajectory taking into account all other vehicle paths to avoid collision and deadlocks. Finally, helyOS dispatches the mission data to the truck for execution (7). This sequence of events is predefined in the helyOS dashboard under a mission recipe named "unload goods", as seen on the right panel of Fig.~\ref{fig:mission_recipe}. This specific sequence is triggered whenever the end-user creates a missions of the  "unload goods" type.

Note that TruckTrix\textsuperscript{\textregistered}-Path and TruckTrix\textsuperscript{\textregistered}-Coop are served from the cloud.  Other kinds of missions would be easily integrated just by adding or rearranging the microservices.

\subsection{Automation of agriculture operations}

\begin{figure}[H]
\centering
\includegraphics[width=0.47\textwidth]{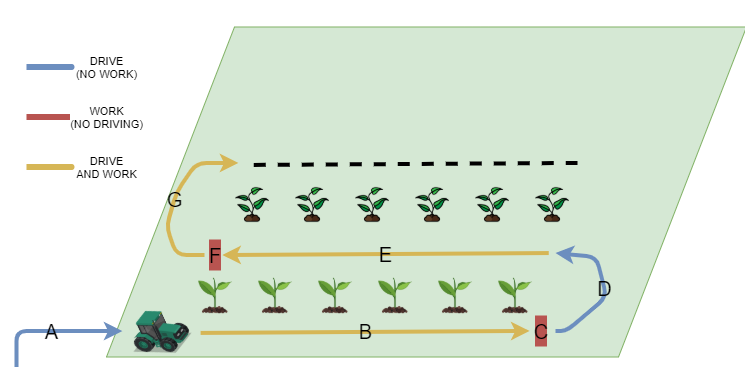}
\caption{Mission example in the Feldschwarm project. In the blue parts of the mission the vehicle just drives, in the yellow parts it drives and simultaneously performs agricultural tasks, in the red parts it performs agricultural tasks only. The parameters for each task can be calculated by one or several microservices. }
\label{fig:cognac}
\end{figure}

 The automation of agricultural processes, known as smart farming, is an important area of yard automation. Smart farming is a challenging concept; it deals with  missions that simultaneously employ several entities of different natures: tractors, drones monitoring field state, weather stations or an external database containing plant data. 

helyOS  simplifies the development and implementation of these systems as has been demonstrated in the  Feldschwarm~\cite{feldschwarm} project, which makes use of the TruckTrix\textsuperscript{\textregistered}-Agriculture motion algorithm. In Fig.~\ref{fig:cognac} one can see a corresponding mission example for the Feldschwarm~\cite{feldschwarm} project, where one mission assignment consists of several tasks. In this project, the helyOS administrator can decide whether the mission data should be calculated for all tasks at once using a complex path planner, or whether the data processing will be distributed over several microservices and the results collected at the end of all calculations. Moreover, a single mission can derive multiple assignments that are addressed to several tractors.

Another example of agriculture automation will be realized as a part of the Cognitive Agriculture (COGNAC) project~\cite{cognac, cognac_osten}. Here, relevant information from an external public database is collected with the help of a connector microservice. This information is used for the creation of tractors missions on the field. For example, one of the assignment planner microservices can use the collected data to estimate the tractor battery consumption before performing a mission. If necessary, it will create intermediate assignments to drive the tractor to a charging station.

\section{Conclusion: applications of helyOS}
\label{sec:concl}

In this paper we presented potential use cases of helyOS and mentioned some demonstrations of applications. The following list summarizes these demonstrations:

\vspace{0.4cm}

\begin{itemize}
    \item In the AutoTruck project~\cite{autotruck},  a fully automated truck was controlled in a logistics yard. 
    
    \item In the SAFE-20 project~\cite{safe20}, we are applying helyOS to a multiple-vehicle logistics yard. The TruckTrix\textsuperscript{\textregistered}-Path and -Coop microservices are used  to guarantee a collision-free trajectory.
    \item The Feldschwarm project~\cite{feldschwarm} presents the work of helyOS in agriculture operations. In this case we created common missions for multiple tractors. These missions include both driving and agriculture operations.  
    \item The COGNAC project~\cite{cognac} for the automation of different field operations needs helyOS to plan single vehicle missions that incorporates the results of different path planners.
     
\end{itemize}

\vspace{0.8cm}

The helyOS architecture proves to be flexible and can be applied to many other use cases by employing appropriated microservices to the HCT and operation modules to the HAC. The development of user interfaces is accelerated by using the HCT as an entry point for web applications. Database customization schemes are promptly available to the front-end by using the GraphQL query language. The current definitions of domains and data flows provide an efficient development guideline for several situations in autonomous driving in delimited areas.

\vspace{1cm}

\printbibliography

\end{document}